\documentclass[aps,prb,twocolumn,superscriptaddress,showpacs,amsmath]{revtex4}

\usepackage{graphicx,amssymb}

\begin{document}

%

\title{Exchange-coupling constants, spin density map, and $Q$ dependence of the
 inelastic neutron scattering intensity in single-molecule magnets}

\author{O. Waldmann}
 \email[Corresponding author.\\E-mail: ]{waldmann@iac.unibe.ch}
 \affiliation{Department of Chemistry and Biochemistry, University of Bern, CH-3012 Bern, Switzerland}

\author{R. Bircher}
 \affiliation{Department of Chemistry and Biochemistry, University of Bern, CH-3012 Bern, Switzerland}

\author{G. Carver}
 \affiliation{Department of Chemistry and Biochemistry, University of Bern, CH-3012 Bern, Switzerland}

\author{A. Sieber}
 \affiliation{Department of Chemistry and Biochemistry, University of Bern, CH-3012 Bern, Switzerland}

\author{H. U. G\"udel}
 \affiliation{Department of Chemistry and Biochemistry, University of Bern, CH-3012 Bern, Switzerland}

\author{H. Mutka}
 \affiliation{Institut Laue-Langevin, 6 rue Jules Horowitz, BP 156, 38042 Grenoble Cedex 9, France}

\date{\today}

\begin{abstract}
The $Q$ dependence of the inelastic neutron scattering (INS) intensity of transitions within the ground-state
spin multiplet of single-molecule magnets (SMMs) is considered. For these transitions, the $Q$ dependence is
related to the spin density map in the ground state, which in turn is governed by the Heisenberg exchange
interactions in the cluster. This provides the possibility to infer the exchange-coupling constants from the
$Q$ dependence of the INS transitions within the spin ground state. The potential of this strategy is
explored for the $M=\pm10\rightarrow\pm9$ transition within the $S=10$ multiplet of the molecule Mn$_{12}$ as
an example. The $Q$ dependence is calculated for powder as well as single-crystal Mn$_{12}$ samples for
various exchange-coupling situations discussed in the literature. The results are compared to literature data
on a powder sample of Mn$_{12}$ and to measurements on an oriented array of about 500 single-crystals of
Mn$_{12}$. The calculated $Q$ dependence exhibits significant variation with the exchange-coupling constants,
in particular for a single-crystal sample, but the experimental findings did not permit an unambiguous
determination. However, although challenging, suitable experiments are within the reach of today's
instruments.
\end{abstract}

\pacs{33.15.Kr, 71.70.-d, 75.10.Jm}

\maketitle

%

\section{Introduction}

An important issue in the field of molecular nanomagnets is the determination of the exchange-coupling
constants between the metal centers in a cluster. In many of these molecules the exchange interactions are in
fact the dominant terms in the spin Hamiltonian, and hence determine the nature of the ground state. The spin
Hamiltonian may be written as
\begin{equation}
\label{H}
 \hat{H} = - \sum_{i\neq j} J_{ij} \hat{\textbf{S}}_i \cdot \hat{\textbf{S}}_j + \hat{H}_A,
\end{equation}
where the first term describes the Heisenberg exchange interactions in the cluster (each exchange bond shall
be counted only once) and the second term the weaker anisotropic contributions (single-ion anisotropy,
anisotropic and antisymmetric exchange, dipole-dipole interactions). The first term will be also denoted as
$\hat{H}_{ex}$.

The single-molecule magnets (SMMs), such as the compound Mn$_{12}$-acetate (or Mn$_{12}$ in short), represent
important examples. The spectacular phenomena observed in these clusters, like the slow relaxation of the
magnetization or quantum tunneling of the magnetization,\cite{Ses93n,Fri96,Tho96,Mn12_Fe8} are intimately
connected to a ground-state spin multiplet, which is characterized by a large spin $S$ and a large anisotropy
splitting of the easy-axis type. Obviously, the topography of the exchange-coupling constants $J_{ij}$ is the
key factor which controls the value of $S$. However, it also has a strong impact on the magnitude of the
anisotropy splitting in the spin ground state. On the one hand, it determines how efficiently the microscopic
anisotropy terms project onto the spin ground state,\cite{Ben90} and on the other hand, it generates
higher-order anisotropy terms via a mixing of spin multiplets (this mechanism is also called $S$
mixing).\cite{Liv02,OW_INS2,Acc06,Hil06}

Typically, the SMMs are modeled by the giant-spin Hamiltonian, which essentially describes the whole molecule
by a single spin $S$. The giant-spin Hamiltonian for Mn$_{12}$, for instance, usually reads (in zero field)
\begin{equation}
 \hat{H}_S = D \left[\hat{S}_z^2 - \frac{1}{3}S(S+1)\right] + B_{40} \hat{O}_4^0(S)
 + B_{44} \hat{O}_4^4(S),
\end{equation}
where $S=10$, and $D = -0.057$~meV, $B_{40} = -2.78 \times 10^{-6}$~meV, and $B_{44} = -3.2 \times
10^{-6}$~meV.\cite{Bir04} The dominating $D$ term splits the spin multiplet into $\pm M$ sublevels, with the
$M = \pm S$ level lowest in energy. It gives rise to a large energy barrier between the $M=+S$ and $M=-S$
states and hence the slow relaxation of the magnetization at low temperatures. The $B_{40}$ and $B_{44}$
terms are fourth-order anisotropy terms.\cite{Mn12_E}

The giant-spin Hamiltonian approach is phenomenological in nature, but allows a very precise (in principle
even exact) description of the energy eigenvalues of the ground-state multiplet, as well as their dependence
on external parameters such as an applied magnetic field.\cite{Ste63} One just has to choose the spin terms
as appropriate (the possible spin terms are restricted by time reversal and cluster symmetry). The results of
basically all experiments on SMMs which focus on the spin ground state, be it magnetization, specific heat,
electron-spin resonance, nuclear magnetic resonance, M\"ossbauer spectroscopy, or else, can be very well
described within this framework.\cite{Mn12_Fe8}

A notable exception is inelastic neutron scattering (INS). The INS selection rule $\Delta M = 0, \pm1$
permits transitions between adjacent $M$ levels, which results in a typical picket-fence INS spectrum with
the $M = \pm S \rightarrow \pm(S-1)$ transition at highest energy transfers.\cite{Cac98,Mir99} While the
giant-spin Hamiltonian works well for the transition energies, the $Q$ dependence of the INS intensity cannot
be described at all in this approach. In fact, the giant-spin Hamiltonian approach predicts the INS intensity
to be constant as function of $Q$,\cite{Cac98,Mir99} while in reality it exhibits a typical oscillatory $Q$
dependence due to the interference effects between the various spin centers in a cluster.\cite{Fur77,OW_INS}
The $Q$ dependence is hence a direct signature of the many-spin nature of the ground-state spin
multiplet.\cite{OW_INS2} The ability of INS to probe the many-spin character of wavefunctions is unrivaled by
other techniques.

This feature of INS suggests the following strategy to infer information on the exchange-coupling constants
in a SMM. The INS intensity is governed by factors $\exp[i\textbf{Q}(\textbf{R}_i-\textbf{R}_j)] \langle
n|\hat{S}_{i\alpha}|m\rangle \langle m|\hat{S}_{j\beta}|n\rangle$, i.e., depends on the geometrical
arrangement of the spin centers in the cluster and the local spin transition matrix elements ($\textbf{Q}$ is
the momentum transfer vector, $\textbf{R}_i$ the position vector of the $i$th spin center, $|n\rangle$ an
eigenstate of the system, and $\alpha,\beta=x,y,z$). The set of elements $\langle
n|\hat{S}_{i\alpha}|m\rangle$ may be called the spin transition map,\cite{OW_INS2} in analogy to the spin
density map, $\langle n|\hat{S}_{i\alpha}|n\rangle$. Via the $\textbf{Q}$ vector in the exponential, the
relative contribution of each of the local matrix elements $\langle n|\hat{S}_{i\alpha}|m\rangle$ to the INS
intensity can be modulated. This gives rise to interference factors and the characteristic $Q$ dependence of
the INS intensity. Hence, the $Q$ dependence allows one to probe the topography of the spin transition map.

Within the space of the ground-state spin multiplet of a SMM, however, the transition matrix elements
$\langle n|\hat{S}_{i\alpha}|m\rangle$ are proportional to the diagonal matrix elements $\langle
n|\hat{S}_{i\alpha}|n\rangle$ in a first approximation, because of the Wigner-Eckart theorem: $\langle \tau S
M|\hat{S}_{i\alpha}| \tau S M' \rangle =
 C^{MM'}_\alpha \times$ $\langle \tau S ||\hat{S}_{i}|| \tau S \rangle = C^{MM'}_\alpha/C^{MM}_\alpha \langle \tau S M|\hat{S}_{i\alpha}| \tau S M
\rangle$, where $\langle \tau S ||\hat{S}_{i}|| \tau S \rangle$ is a reduced matrix element and
$C^{MM'}_\alpha$ are constants (essentially Wigner-3$j$ symbols). In other words, the $Q$ dependence of an
INS transition within the ground-state multiplet is related to the spin density map, which in turn, as
evident already from physical intuition, is determined by the exchange-coupling constants in the cluster.
Hence, the $Q$ dependence of the INS transitions between the $\pm M$ levels of the spin ground state of a SMM
should allow one to retrieve the exchange-coupling constants.

The novel concept in this approach should be noted. Usually, the coupling constants are determined by
probing, either via spectroscopic techniques or thermal excitation, the energies of the higher lying spin
multiplets. The method suggested here, in contrast, retrieves the information exclusively from the
ground-state multiplet. This could be, e.g., an advantage in cases where the higher lying spin multiplets are
not accessible to experiment for one or the other reason.

This work aims at exploring this strategy with the example of Mn$_{12}$. The $Q$ dependence is calculated for
various exchange-coupling topographies considered before in the literature, and compared to experiment. The
two situations of a powder and a single-crystal sample of Mn$_{12}$ are considered. In the powder case, the
calculations are compared to the experimental data of Hennion {\it et al.};\cite{Hen97} for the
single-crystal case data has been recorded on a quasi-single-crystal sample, which consisted of an oriented
array of about 500 single crystals of Mn$_{12}$.

The outline of the manuscript is as follows. In the next section the relevant theoretical basics are
presented. In Sec.~III some exchange-coupling topographies are described and the reduced matrix elements or
projection coefficients, respectively, calculated. In Sect.~IV experimental details are given. Sections~V and
VI discuss the $Q$ dependence of a powder and single-crystal SMM sample, respectively. Section~VII finally
presents a conclusion. Some supporting information is provided in the Appendix.

%

\section{Basics}

The differential neutron scattering cross section for spin clusters is given by \cite{Fur79,Gue85}
\begin{equation}
\label{cross_section}
 \frac{d^2 \sigma}{d\Omega d\omega} = C(Q,T) \sum_{nm} \frac{e^{-\beta E_n}}{Z(T)}
 I_{nm}(\textbf{Q}) \delta\left(\omega-\frac{E_m-E_n}{\hbar}\right)
\end{equation}
with
\begin{eqnarray}
\label{I_nm}
 I_{nm}(\textbf{Q}) &=& \sum_{ij} F_i(Q) F_j(Q) e^{i\textbf{Q}\cdot\textbf{R}_{ij}}
 \sum_{\alpha\beta} ( \delta_{\alpha\beta}-\frac{Q_{\alpha}Q_{\beta}}{Q^2} )
 \cr&&\times
  \langle n|\hat{S}_{i\alpha}|m\rangle \langle m|\hat{S}_{j\beta}|n\rangle.
\end{eqnarray}
Here, $C(Q,T)=(\gamma e^2/m_e c^2) (k'/k) \exp[-2 W(Q,T)]$ (all symbols have the usual meaning), $\beta =
1/(k_B T)$, $Z(T)$ is the partition function, $F_i(Q)$ is the magnetic form factor of the $i$th spin center,
$\textbf{Q} = \textbf{k} - \textbf{k}'$ is the transferred momentum, and
$\textbf{R}_{ij}=\textbf{R}_i-\textbf{R}_j$ is the distance vector between spin $i$ and $j$. $|n\rangle$
denotes an eigenstate with energy $E_n$ of the microscopic spin Hamiltonian $\hat{H}$. In this work,
$|n\rangle$ is one of the $2S+1$ states of the ground-state multiplet of a SMM. In the following, also the
abbreviation $l_{\alpha} = Q_{\alpha} / Q$ is used.

For a powder sample in zero magnetic field the INS intensity is obtained by averaging over all orientations
of $\textbf{Q}$, which yields \cite{OW_INS}
\begin{eqnarray}
\label{I_nm_powder}
 I_{nm}(Q) &=& \sum_{ij} F^*_i(Q) F_j(Q) \{ \frac{2}{3} j_0(Q R_{ij}) {\bf \tilde{S}}_i \cdot {\bf \tilde{S}}_j
 \cr &&
  + j_2(Q R_{ij}) \sum_q T^{(2)*}_q({\bf R}_{ij}) T^{(2)}_q( {\bf \tilde{S}}_i{\bf \tilde{S}}_j) \},
\end{eqnarray}
where $j_k$ is the spherical Bessel function of order $k$, $T^{(k)}_q({\bf v})$ is the $q$th component of the
spherical tensor of rank $k$ constructed from the Cartesian vector ${\bf v}$, and $T^{(2)}_q( {\bf
\tilde{S}}_i{\bf \tilde{S}}_j)$ represents the tensor product $[T^{(1)}({\bf \tilde{S}}_i) \otimes
T^{(1)}({\bf \tilde{S}}_j)]^{(2)}_q$. The ordered products $\tilde{S}_{i \alpha} \tilde{S}_{j \beta}$, which
appear in the explicit expression of $T^{(2)}_q( {\bf \tilde{S}}_i{\bf \tilde{S}}_j)$, stand for $\langle
n|\hat{S}_{i \alpha}|m\rangle\langle m| \hat{S}_{j \beta}|n\rangle$. Equation~(\ref{I_nm_powder}) can be also
written more compactly as \cite{OW_CsFe8_INS2}
\begin{eqnarray}
\label{I_nm_powder2}
 I_{nm}(Q) = \sum_{ij} \sum_{kq} f^{kq}_{ij}(Q,{\bf R}_{ij}) U^{(k)}_q({\bf
\tilde{S}}_i {\bf \tilde{S}}_j),
\end{eqnarray}
with the interference factors $f^{kq}_{ij}(Q,{\bf R}_{ij})$ ($k = 0, 2$ and $|q| \leq k$) and the symmetrized
spherical tensors $U^{(k)}_q$ (which are proportional to Re[$T^{(k)}_q$] for $q \geq 0$ and Im[$T^{(k)}_q$]
for $q < 0$). Explicit expressions for Eqs.~(\ref{I_nm_powder}) and (\ref{I_nm_powder2}) are given in
Refs.~[\onlinecite{OW_INS2}] and [\onlinecite{OW_CsFe8_INS2}], respectively.

Often it is desirable to describe the low-lying excitations of a spin cluster by an effective spin
Hamiltonian $\hat{\bar{H}}$ instead of the microscopic spin Hamiltonian $\hat{H}$. The eigenstates of the
effective spin Hamiltonian will be denoted as $|\bar{n}\rangle$, and the spin operators acting in the space
of the effective spin Hamiltonian as $\hat{\bar{S}}_{j\beta}$ (effective spin operators). Since
$\hat{\bar{H}}$ is supposed to yield the same energies as $\hat{H}$, a distinction between $E_n$ and
$\bar{E}_n$ is not necessary. For a more detailed description of the relationships between effective and
microscopic spin Hamiltonian see e.g. Ref.~[\onlinecite{OW_INS2}]. Using the effective spin Hamiltonian, the
$Q$ dependence of the INS intensity can be reproduced with a high degree of accuracy by a first-order
perturbation theory approach.\cite{OW_CsFe8_INS2,OW_INS2}

In the present context of a SMM, where the excitations within the ground-state spin multiplet are of primary
interest, the effective spin Hamiltonian $\hat{\bar{H}}$ simply corresponds to the well-known giant-spin
Hamiltonian $\hat{H}_S$. The effective spin operators $\hat{\bar{S}}_{j\beta}$ are then related to the
components of the total spin operator $\hat{S}_\beta$ via the projection coefficients $\Gamma_1(i)$, i.e.,
$\hat{\bar{S}}_{i\alpha} = \Gamma_1(i) \hat{S}_{\alpha}$.\cite{OW_INS2} The matrix elements of the local spin
operators $\hat{S}_{i\alpha}$ hence simply become
\begin{equation}
\label{S_eff}
 \langle n|\hat{S}_{i\alpha}|m\rangle = \Gamma_1(i) \langle \bar{n}|\hat{S}_{\alpha}|\bar{m}\rangle.
\end{equation}
The eigenstates $|\bar{n}\rangle$ are obtained from the diagonalization of the giant-spin Hamiltoinan in the
spin space $|SM\rangle$. The projection coefficients are determined as usual by\cite{Ben90}
\begin{equation}
\label{gamma}
 \Gamma_1(i) = \frac{\langle\tau S||T^{(1)}(S_i)||\tau S\rangle}{\langle S||T^{(1)}(S)||S\rangle},
\end{equation}
where $\langle\tau S||T^{(1)}(S_i)||\tau S\rangle$ is a reduced matrix element related to the eigenstate
$|\tau SM\rangle$ of the Heisenberg-exchange part $\hat{H}_{ex}$ of the microscopic spin Hamiltonian, and
$\langle S||T^{(1)}(S)||S\rangle = \sqrt{(2S+1)S(S+1)}$ is the reduced matrix element in the spin space
$|SM\rangle$.

A subtle point shall be mentioned here. The described procedure essentially correspond to the strong-exchange
limit,\cite{Ben90} i.e., the first-order perturbational treatment of the magnetic anisotropy. However, the
strong-exchange limit does not yield all relevant spin terms in the giant-spin Hamiltonian correctly, such as
for instance the fourth-order terms, because it neglects a mixing of the spin
multiplets.\cite{Liv02,OW_INS2,Acc06,Hil06} In principle, a similar statement is also true for the effective
spin operators.\cite{OW_INS2} However, it turns out that the mixing of the spin multiplets has a small effect
on the topography of the matrix elements $\langle\bar{n}|\hat{S}_{i\alpha}|\bar{m}\rangle$ even for rather
large magnetic anisotropy, such that the $Q$ dependence of the INS intensity is very well reproduced in first
order. This has been explicitly demonstrated in particular for the Mn$_{12}$ cluster, see Fig.~3 of
Ref.~[\onlinecite{OW_INS2}]. Therefore one finds that the local spin operators are related to the total spin
operator via factors (the projection coefficients), which are determined completely by the Heisenberg
interactions in the system.

Insertion of Eq.~(\ref{S_eff}) into the INS formula (\ref{I_nm}), as appropriate for a single-crystal
material, yields
\begin{equation}
\label{I_nm_S}
 I_{nm}({\bf Q}) = \mathcal{F}({\bf Q})
  \sum_{\alpha\beta} ( \delta_{\alpha\beta}-l_{\alpha}l_{\beta} )
  \mathrm{Re}\!\!\left[\langle \bar{n}|\hat{S}_{\alpha}|\bar{m}\rangle \langle \bar{m}|\hat{S}_{\beta}|\bar{n}\rangle\right]
\end{equation}
with the interference factor
\begin{equation}
\label{F_S}
 \mathcal{F}({\bf Q}) = \sum_{ij} F_i(Q) F_j(Q) \cos({\bf Q}\cdot{\bf R}_{ij}) \Gamma_1(i)
\Gamma_1(j).
\end{equation}
Because of the symmetry with respect to the interchange $ij \rightarrow ji$, the exponential is replace by
the cosine and the real part, Re$[]$, is introduced. It is interesting to note that the interference effects,
described by $\mathcal{F}({\bf Q})$, factorize.

For a powder sample, Eq.~(\ref{S_eff}) should be inserted into, for instance, Eq.~(\ref{I_nm_powder2}), which
effectively replaces  $U^{(k)}_q({\bf \tilde{S}}_i {\bf \tilde{S}}_j)$ by $\Gamma_1(i) \Gamma_1(j)
U^{(k)}_q({\bf \tilde{S}}{\bf \tilde{S}})$. The result is
\begin{equation}
\label{I_nm_powder2_S}
 I_{nm}(Q) = \sum_{kq} \mathcal{F}^{kq}(Q) U^{(k)}_q({\bf \tilde{S}}{\bf \tilde{S}})
\end{equation}
with the interference factors
\begin{equation}
\label{F_powder S}
 \mathcal{F}^{kq}(Q) = \sum_{ij} f^{kq}_{ij}(Q,{\bf R}_{ij}) \Gamma_1(i)\Gamma_1(j).
\end{equation}
In contrast to the single-crystal case, where the interference factor depends on the orientation of the
transferred momentum vector, here the interference factors depend only on its magnitude. There is an
instructive relation between Eqs.~(\ref{F_S}) and (\ref{F_powder S}), which is outlined in the Appendix.

The giant-spin approach has been often used to analyze INS data on SMMs.\cite{Cac98,Mir99} However, as
discussed in the introduction, it disregards the many-spin nature of the cluster; the interference factors
are hence constants, and the predicted INS $Q$ dependencies just flat. This fundamental deficiency is
overcome by Eqs.~(\ref{F_S}) and (\ref{F_powder S}), which reproduce $Q$ dependencies accurately. They hence
are much more preferable than the giant-spin approach. Importantly, for both the powder and single-crystal
case the interference factors depend solely on the geometrical arrangement of the spin centers and the
projection coefficients on each site. They do not depend, for instance, on the particular form of the
giant-spin Hamiltonian. Hence, as the structure of the cluster is known, the measured $Q$ dependencies
provide direct information on the projection coefficients, and wherewith on the exchange-coupling constants.
This is the basic idea of the present work, and Eqs.~(\ref{F_S}) and (\ref{F_powder S}) are the explicit
formulations of it.

%

\section{Projection Coefficients}

\begin{figure}
\includegraphics{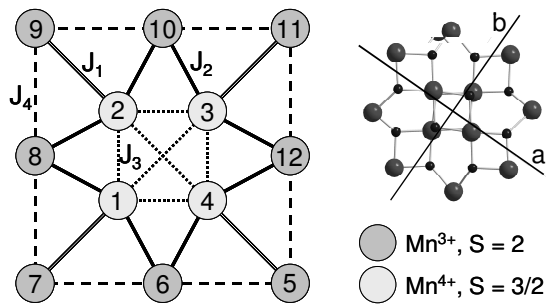}
\caption{\label{fig1} Schematic representation of the spin centers and exchange-coupling paths in Mn$_{12}$.
The spins 1-4 are denoted as core and the spins 5-12 as the crown. The right graphic shows the molecular
structure of Mn$_{12}$ viewed along the crystallographic $c$ axis (only the Mn and the bridging oxygen atoms
are shown). The orientations of the crystallographic $a$ and $b$ axes of the tetragonal unit cell are
indicated. The core and crown are rotated with respect to the $a$ and $b$ axes such that the axis through
spins 1 and 3 is rotated by 12.33$^\circ$ and that through spins 7 and 11 by 10.91$^\circ$.}
\end{figure}

As shown in the previous section, the evaluation of the INS $Q$ dependence involves the geometrical structure
of the cluster and the projection coefficients. The molecular structure of Mn$_{12}$ is precisely know from
x-ray crystallography (Mn$_{12}$ crystallizes in space group I$_4$ and exhibits a molecular S$_4$ symmetry
axis, see also Fig.~\ref{fig1}).\cite{Lis80} In this section, the projection coefficients for Mn$_{12}$ will
be determined for various exchange-coupling topographies considered before in the literature.

The exchange-coupling graph appropriate for Mn$_{12}$ is presented in Fig.~\ref{fig1}. It involves the four
exchange constants $J_1$, $J_2$, $J_3$, and $J_4$. Due to the S$_4$ symmetry of the cluster, the projection
coefficients assume three different values, namely
\begin{eqnarray}
\label{gamma_x}
 \Gamma_1(A) &\equiv& \Gamma_1(1) = \Gamma_1(2) = \Gamma_1(3) = \Gamma_1(4),\\
 \Gamma_1(B) &\equiv& \Gamma_1(6) = \Gamma_1(8) = \Gamma_1(10) = \Gamma_1(12),\\
 \Gamma_1(C) &\equiv& \Gamma_1(5) = \Gamma_1(7) = \Gamma_1(9) = \Gamma_1(11).
\end{eqnarray}
The indices $A$, $B$, and $C$ will be used also to notate the three sublattices consisting of the spins
$\{1,2,3,4\}$, $\{6,8,10,12\}$, and $\{5,7,9,11\}$, respectively. Since $\Gamma_1(A) + \Gamma_1(B) +
\Gamma_1(C) = 1/4$ the projection coefficients are not independent of each other, and the $Q$ dependence of
the INS intensity is effectively governed by only two independent parameters.

The first idea on the coupling situation in Mn$_{12}$ came from magneto-chemical considerations, which
suggested that $J_1$ is antiferromagnetic and much larger than the other three coupling
constants.\cite{Ses93} Hence, the spins of e.g. the Mn ions 1 and 7 couple to an overall spin of 1/2, and
similar for the spins 2\&9, 3\&11, and 4\&5. In order to arrive at a $S=10$ ground state, these spins are
then coupled ferromagnetically to the remaining four Mn(III) spins 6, 8, 10, and 12. Following
Ref.~[\onlinecite{Har96}], this scheme will be called the "Florentine" coupling scheme. Since the
ground-state spin wavefunction is constructed here by a successive coupling of spins, the projection
coefficients can be calculated easily using the Wigner-Eckart theorem and Racah formalism.\cite{Ben90}

A second scheme consists of coupling the eight Mn(III) spins ferromagnetically to yield an overall spin
$S_{BC} = 16$, to couple the four Mn(IV) spins ferromagnetically to yield a spin $S_A = 6$, and then to
couple these two spins antiferromagnetically to yield $S = S_{BC} - S_A = 10$.\cite{Har96} This coupling
scheme is based on the rational that $J_1$ and $J_2$ are both antiferromagnetic, while $J_3$ and $J_4$ are
assumed to be ferromagnetic. It will be called the "A+(B+C)" scheme. The projection coefficients again can be
calculated easily using the Wigner-Eckart theorem and Racah formalism.

A third scheme consists of coupling the four spins of each sublattice $A$, $B$, $C$ ferromagnetically, then
to couple the spins of sublattices $A$ and $B$ antiferromagnetically to yield $S_{AB} = 4$, and to finally
couple this spin with that of sublattice $C$ antiferromagnetically to yield $S= S_{AB} - S_C = 10$. This
coupling scheme is based on the rational that $J_1$ and $J_2$ are both antiferromagnetic and dominating, such
that the frustration induced by $J_3$ and $J_4$ is negligible. It will be called "(A+B)+C". The calculation
of the projection coefficients proceeds as for the previous two schemes.

The above three schemes approximate the wavefunction of the ground-state spin multiplet by a spin
wavefunction $|\nu SM\rangle$, where $\nu$ denotes intermediate spin quantum numbers. The associated
spin-coupling schemes were suggested by the different assumptions concerning the exchange-coupling
topography. A more penetrating approach is of course to specify the values of the coupling constants $J_1,
\ldots, J_4$ and to calculate the ground-state wavefunction numerically by exact diagonalization. Such an
approach has been pursued for Mn$_{12}$ by three groups. Raghu {\it et al.} suggested the set $J_1$ =
$-$215~K, $J_2$ = $J_3$ = $-$85~K, $J_4$ = 64.5~K.\cite{Rag01} The projection coefficients can be determined
from the spin densities given in their work. Regnault {\it et al.} deduced the values $J_1$ = $-$119~K, $J_2$
= $-$118~K, $J_3$ = 8~K, $J_4$ = $-$23~K.\cite{Reg02} No information is given which would allow a
determination of the projection coefficients. Hence we calculated them numerically using a sparse-matrix
iterative subspace technique.\cite{OW_SPINDYN,OW_INS2} Finally, Chaboussant {\it et al.} deduced the values
$J_1$ = $-$67.2~K, $J_2$ = $-$61.8~K, $J_3$ = $-$7.8~K, $J_4$ = $-$5.6~K.\cite{Cha04,Hon05} Here too the
projection coefficients were calculated by us numerically. The exchange-coupling constants were also
evaluated by LDA+U and DFT calculations,\cite{Bou02,Par04} but these additional sets do not change the
conclusions of this work and are hence not considered. The works Refs.~[\onlinecite{Kat99,Zve96,Tup02}],
which start from an effective 8-spin model for Mn$_{12}$, are mentioned for completeness.

\begin{table}
\caption{\label{tab1} Projection coefficients for the exchange-coupling topographies discussed in the text.}
\begin{ruledtabular}
\begin{tabular}{cccc}
 scheme & $\Gamma_1(A)$ & $\Gamma_1(B)$ & $\Gamma_1(C)$  \\
\hline
Florentine               & $-$0.05 & 0.2 & 0.1 \\
A+(B+C)            & $-$0.13636 & 0.19318 & 0.19318 \\
(A+B)+C            & $-$0.1 & 0.15 & 0.2 \\
Raghu {\it et al.}       & $-$0.09019 & 0.18103 & 0.15916 \\
Regnault {\it et al.}    & $-$0.12472 & 0.18376 & 0.19096 \\
Chaboussant {\it et al.} & $-$0.12121 & 0.18398 & 0.18723
\end{tabular}
\end{ruledtabular}
\end{table}

The projection coefficients of the six exchange-coupling topographies just described are compiled in
Table~\ref{tab1}. For all cases, $\Gamma_1(A)$ is negative while $\Gamma_1(B)$ and $\Gamma_1(C)$ are
positive. This reflects the antiferromagnetic alignment of the spins of the Mn(IV) core with respect to the
spins on the Mn(III) crown. The different exchange-coupling topographies clearly express themselves in
different projection coefficients. However, the variation is not very pronounced, with the exception of the
Florentine coupling scheme. For this scheme $\Gamma_1(A)$ and $\Gamma_1(C)$ are significantly smaller in
magnitude, which reflects the assumption of a dominant $J_1$ in this scheme.

%

\section{Experiments}

In the next sections, the $Q$ dependence of the INS transition between the $M=\pm10$ and $M=\pm9$ levels of
the $S=10$ ground-state multiplet of Mn$_{12}$ will be considered. It occurs at an energy transfer of
1.24~meV, and is the strongest inelastic peak at liquid-He temperatures.

The best measurement so far of the $Q$ dependence of this transition in powder samples of Mn$_{12}$ has been
reported by Hennion {\it et al.}.\cite{Hen97} These authors studied a partially deuterated Mn$_{12}$ sample
on the triple-axis spectrometers 4$F$1 and 1$T$ at the Laboratoire Leon Brillouin, Saclay, France, at a
temperature of 1.55~K. The calculation of the $Q$ dependence for powder Mn$_{12}$, which will be presented in
the next section, hence will be compared to this data set.

In order to determine the $Q$ dependence of a single-crystal sample, we performed measurements on an oriented
array of about 500 non-deuterated single crystals of Mn$_{12}$. The crystals were synthesized following
literature procedures.\cite{Lis80} The dimensions of the needle-shaped crystals were approximately
5$\times$0.5$\times$0.5~mm$^3$, and the masses ca. 2~mg. Mn$_{12}$ crystallizes in the space group
I$\bar{4}$, hence the needle axis coincides with the magnetic anisotropy axis $z$. The oriented array of
single crystals was obtained by placing crystals in long, narrow grooves milled into aluminium platelets,
such that the needle ($=z$) axes were aligned along the grooves.\cite{OW_TRINS} A total of 15 such platelets
which each held about 30$-$35 single crystals were stacked in an aluminium container and sealed therein. The
INS intensity was measured on the time-of-flight spectrometer IN5 at the Institut Laue-Langevin, Grenoble,
France. The initial neutron wavelength was set to 5.9~{\AA}; the resolution at the elastic line was
60~$\mu$eV. The sample was inserted into an orange cryostat, permitting a sample temperature of 1.5~K. The
data were corrected for the contribution of the sample container and the detector efficiency using a vanadium
standard.

The alignment of the $z$ axis of the quasi-single-crystal sample with respect to the transferred momentum
vector $\textbf{Q}$ crucially influences the measured $Q$ dependence. In simulations, details of the
experiment have hence to be considered. The laboratory frame $XYZ$ is chosen such that the $X$ axis points
along the wave vector of the incident horizontal neutrons, $\textbf{e}_X = \textbf{k}/k$, and the $Z$ axis
into the sky. In the experimental configuration of the IN5 spectrometer, the $Y$ axis is then directed
towards the side of the detector banks. The $\textbf{Q}$ vector lies in the $XY$ scattering plane and is
described by its magnitude $Q$ and angle $\varphi_Q$, ${\bf Q} = Q ( \cos \varphi_Q,\sin \varphi_Q,0)$, with
$Q_Y \leq 0$. The calculation of matrix elements, however, is most conveniently done in the $xyz$ frame of
the molecular magnetic axes. As will be shown in the next section, the cluster can be treated as uniaxial,
and the orientation of the sample hence described by the polar angles $\theta$ and $\varphi$ of its $z$ axis
in the $XYZ$ frame. This procedure is not exactly correct, because the cluster coordinates are not invariant
under rotations around $z$. We found, however, that this subtlety does not affect the conclusions of this
work and hence do not exploit it further here. In our calculations the $x$, $y$ axes were along the
crystallographic $a$, $b$ directions, see Fig.~\ref{fig1}. The $z$ axis in the $XYZ$ frame is generated by
two rotations, ${\bf z} = R_Z(\varphi) R_Y(\theta) {\bf Z}$. The transformation of a vector from the $XYZ$ to
the $xyz$ frame is thus $R_Y^{-1}(\theta) R_Z^{-1}(\varphi)$, and for $\textbf{Q}$ in the $xyz$ frame one
obtains ${\bf Q} = Q ( \cos\theta \cos(\varphi_Q-\varphi), \sin(\varphi_Q-\varphi), \sin\theta
\cos(\varphi_Q-\varphi))$. The experimental data was taken for four different configurations, namely with the
$z$ axis along $(0,0,1)$, $(1,0,0)$, $(0,1,0)$, and $(-1,1,0)$, where $(X,Y,Z)$ denotes a vector in the $XYZ$
frame.

For comparison of experiment and calculation, it also has to be considered that the IN5 spectrometer measures
the intensity as function of the direction $\textbf{k}'/k'$, and not $\textbf{Q}$. The data analysis software
LAMP,\cite{lamp} however, which was used for data reduction, yields the intensity as function of $Q$ (which
is related to $\varphi_Q$ via the scattering triangle). Hence, the experimental $Q$ dependencies include an
implicit variation of $\varphi_Q$, i.e., the curves do not correspond to $Q$ scans as they would be obtained
on, e.g., triple-axis instruments.

%

\section{$Q$ dependence of a powder SMM sample}

In this section, the $Q$ dependence of the $M=\pm10 \rightarrow \pm9$ transition of a powder sample of
Mn$_{12}$ will be calculated for the six exchange topographies described in Sec.~III, and compared to the
experimental results of Hennion {\it et al.}.\cite{Hen97} In order to proceed, the matrix elements $\langle
\bar{n}|\hat{S}_{\alpha}|\bar{m}\rangle$ have to be evaluated. Here $|\bar{n}\rangle$ denotes one of the two
states of the $M=\pm10$ level, and $|\bar{m}\rangle$ one of the two states of the excited $M=\pm9$ level (the
difference between $|n\rangle$ and $|\bar{n}\rangle$ is recalled, Sec.~II).

Some points need clarification. First, a zero-field situation is considered. Hence, the wavefunctions may be
chosen as real. The $M=\pm10$ and $M=\pm9$ levels each consist of two states, which shall be denoted as
$|0\rangle_\pm$ and $|1\rangle_{\pm}$, respectively. Since for these levels the tunnel splitting is way to
small to be observed in INS, the intensity of the $M=\pm10 \rightarrow \pm9$ transition in fact originates
from four transitions. This is easily accounted for in the equations by considering each square of matrix
elements to correspond to the sum $\sum_{\mu\nu=\pm} \langle 0|_{\mu}\hat{S}_{\alpha}|1\rangle_{\nu}\langle
1|_{\nu}\hat{S}_{\beta}|0\rangle_{\mu}$. Second, the eigenstates of the giant-spin Hamiltonian are not
eigenstates of $S_z$ because of the presence of terms such as $B_{44}\hat{O}_4^4$.\cite{Mn12_E} They hence
cannot be labeled by $M$ (we nevertheless use the notion of, e.g., a $M=\pm10$ level, its correct meaning
should be obvious). The four lowest-lying states in fact are very well described by
$|0\rangle_{\pm}=(|10\rangle\pm|-10\rangle)/\sqrt2$ and $|1\rangle_{\pm}=(|9\rangle\pm|-9\rangle)/\sqrt2$.
This means (i) that the matrix elements for the operator $\hat{S}_z$ are essentially zero, and (ii) that the
Mn$_{12}$ molecule can be considered as magnetically uniaxial (in fact $\sum_{\mu\nu=\pm} \langle
0|_{\mu}\hat{S}_{\alpha}|1\rangle_{\nu}\langle 1|_{\nu}\hat{S}_{\beta}|0\rangle_{\mu} =
\sum_{M=\pm10,M'=\pm9} \langle M|\hat{S}_{\alpha}|M'\rangle \langle M'|\hat{S}_{\beta}|M\rangle$). These
considerations are generally valid for the states of a SMM at the bottom of the energy barrier (in zero
magnetic field).

With the choice of real wavefunctions, $\langle \bar{n}|\hat{S}_z|\bar{m}\rangle = 0$, and uniaxiality, only
$U^{(0)}_0$ and $U^{(2)}_0$ in Eq.~(\ref{I_nm_powder2}) are non-zero with $U^{(2)}_0 = -1/\sqrt6 U^{(0)}_0$.
Hence one obtains
\begin{equation}
 \label{I_powder SMM}
 I_{nm}(Q) = \frac{2}{3} U^{(0)}_0 \mathcal{F}_{p}(Q)
\end{equation}
with the interference factor
\begin{eqnarray}
\label{F_powder SMM}
 \mathcal{F}_{p}(Q) &=& \sum_{ij} F_i(Q)F_j(Q) \Gamma_1(i)\Gamma_1(j) \cr && \times
 [j_0(QR_{ij})- j_2(QR_{ij}) \frac{3R^2_{ij,z}-R^2_{ij}}{4R^2_{ij}} ].
\end{eqnarray}
This is a rather general result. It holds for the INS transitions in a powdered SMM at the bottom of the
energy barrier [it also implies that the $Q$ dependence does not depend on $M$, only for states near the top
of the barrier it may display deviations from $\mathcal{F}_{p}(Q)$ due to the effects of $\hat{S}_z$].
Interestingly, the interference factor again factorizes. The single-spin approach yields $I_{nm}(Q) \propto
U^{(0)}_0$, i.e., $\mathcal{F}_{p}(Q)$ = const.\cite{Cac98,Mir99,OW_INS2} Comparison with Eq.~(\ref{I_powder
SMM}) hence again nicely demonstrates the effect of the many-spin nature of a real spin cluster, which is
expressed here by the interference factor $\mathcal{F}_{p}(Q)$.

\begin{figure}
\includegraphics{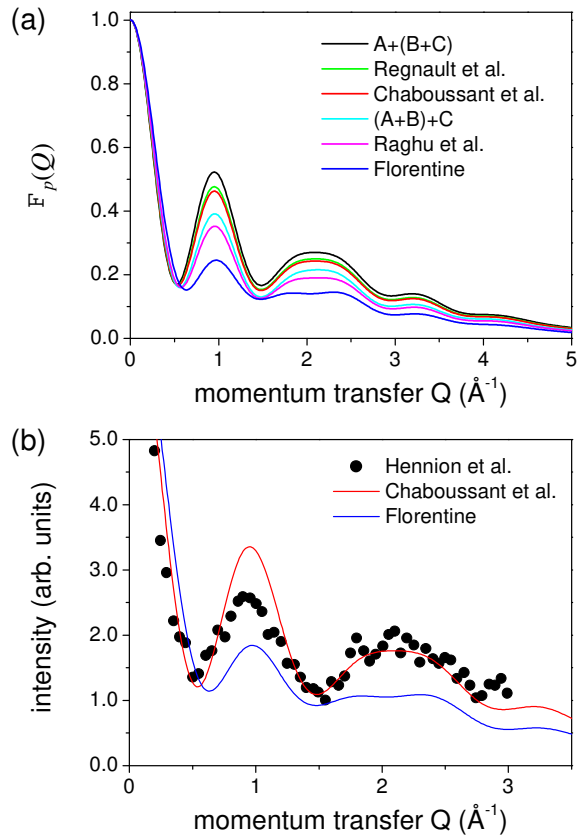}
\caption{\label{fig2}(Color online) (a) Calculated $Q$ dependencies of the intensity of the $M=\pm10
\rightarrow \pm9$ transition in a powder sample of Mn$_{12}$ for the six exchange-coupling topographies
discussed in the text. Curves are labeled in the order of the height of the maximum at 1~{\AA}$^{-1}$. (b)
Experimental data of Hennion {\it et al.} (reproduced from Ref.~[\onlinecite{Hen97}]). For comparison also
the calculated curves for the Chaboussant {\it et al.} and Florentine coupling schemes are reproduced (the
Chaboussant {\it et al.} curve was scaled such as to reproduce the data, the Florentine curve was then scaled
by the same factor).}
\end{figure}

The $Q$ dependencies calculated from Eq.~(\ref{F_powder SMM}) are shown in Fig.~\ref{fig2}(a) for the
projection coefficients given in Table~I. The form factors $F_i(Q)$ were estimated with the standard
analytical approximations.\cite{FFILL} The non-zero value for $Q \rightarrow 0$ is typical for a $\Delta S =
0$ transition.\cite{Gue85} After a steep drop with a minimum at about 0.5~{\AA}$^{-1}$, the curves exhibit a
first pronounced maximum at about 1.0~{\AA}$^{-1}$, display a broad feature between ca. 1.5 and
2.8~{\AA}$^{-1}$, followed by further oscillations at higher $Q$ values. Apparently, although significant
variation is seen, the $Q$ dependence of powder samples does not depend very sensitively on the
exchange-coupling topography. The most significant effect is displayed by the height of the first maximum,
which is most reasonably measured with respect to the two minima to the left and right. A qualitative
comparison of Fig.~\ref{fig2}(a) with Table~I reveals that the height of the first maximum is correlated to
the magnitude of $\Gamma_1(A)$. However, presumably it would require very precise experiments to measure this
height reliably. The broad feature in the range 1.5$-$2.8~{\AA}$^{-1}$ looks very similar for all considered
cases, with the exception of the Florentine coupling scheme, for which it exhibits two rather well resolved
maxima at about 1.8 and 2.4~{\AA}$^{-1}$.

Figure~\ref{fig2}(b) shows the experimental data from Ref.~\onlinecite{Hen97} for comparison. All calculated
$Q$ dependencies reproduce the experimental curve more or less (the Chaboussant \emph{et al.} curve has been
also drawn as a representative for the five schemes other than the Florentine scheme, which all give similar
agreement). The Florentine coupling scheme, however, shows the least agreement, see Fig.~\ref{fig2}(b).
First, the experimental height of the first maximum is clearly larger than calculated (in particular if one
considers that e.g. incoherent scattering due to hydrogens should reduce the experimentally observed height).
Second, the broad feature seems to be better reproduced by the other curves.

Hence, as a summary, the powder $Q$ dependence shows some variation for different exchange-coupling
topographies, but they are not easily differentiated in the experiment. However, it is fair to conclude that
the Florentine coupling scheme is not consistent with the data, and can be ruled out (as has been found also
before\cite{Rag01,Reg02,Cha04,Hon05}). Hence, the powder $Q$ dependence has some potential for
differentiating between very different coupling topographies (it should be recalled that the Florentine
scheme is distinguished from the others by its very dominant $J_1$).

%

\section{$Q$ dependence of a single-crystal SMM sample}

In this section, the $Q$ dependence of the $M=\pm10 \rightarrow \pm9$ transition of a single-crystal sample
of Mn$_{12}$ will be considered. In this case, the INS intensity does not only depend on the magnitude of
$\textbf{Q}$, as in the powder case, but also on its orientation with respect to the sample. The orientation
of $\textbf{Q}$ enters both Eqs.~(\ref{I_nm_S}) and (\ref{F_S}). Equation~(\ref{I_nm_S}), however, can be
further simplified. With real wavefunctions, the matrix elements of $\hat{S}_x$ and $\hat{S}_z$ are real,
while those of $\hat{S}_y$ are imaginary. Hence the sum over $\alpha$, $\beta$ reduces to
\begin{eqnarray}
\label{I_nm_SMM}
 I_{nm}({\bf Q}) &=& \mathcal{F}({\bf Q}) [\sum_{\alpha}  ( 1 - l^2_\alpha ) |\langle \bar{n}|\hat{S}_\alpha|\bar{m}\rangle|^2
  -2 l_x l_z
 \cr &&\times
  \langle \bar{n}|\hat{S}_x|\bar{m}\rangle \langle \bar{n}|\hat{S}_z|\bar{m}\rangle ].
\end{eqnarray}
Further considering that the matrix elements of $\hat{S}_z$ are basically zero and that the cluster can be
treated as magnetically uniaxial, as discussed in section~V, one arrives at the approximation $I_{nm}({\bf
Q}) = \mathcal{F}({\bf Q})( 2 - l^2_x - l^2_y ) |\langle \bar{n}|\hat{S}_x|\bar{m}\rangle|^2$.

\begin{figure}
\includegraphics{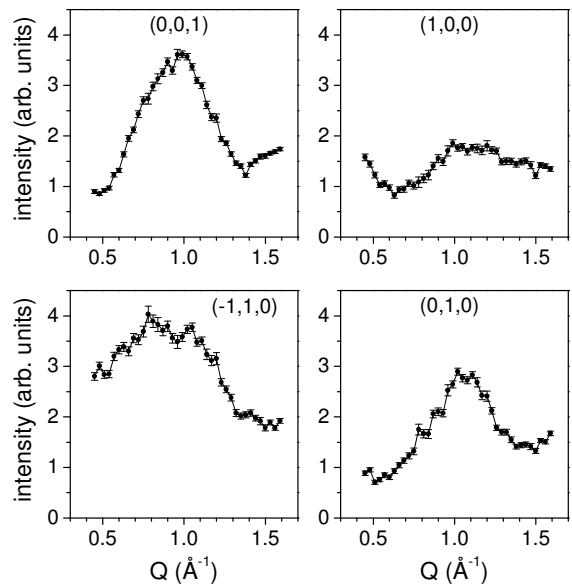}
\caption{\label{fig3} Measured $Q$ dependencies of the intensity of the $M=\pm10 \rightarrow \pm9$ transition
in a quasi-single-crystal sample of Mn$_{12}$ for the four orientations of the $z$ axis indicated in each
panel. The normalization is arbitrary but identical for each curve; the curves hence correctly reflect
relative intensities.}
\end{figure}

\begin{figure}
\includegraphics{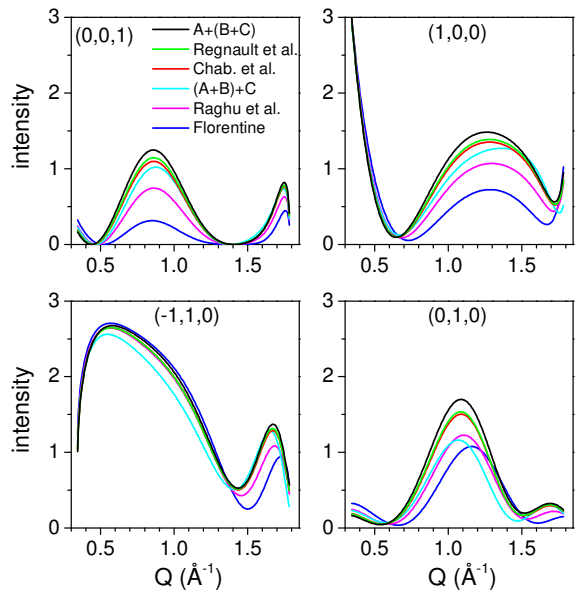}
\caption{\label{fig4}(Color online) Calculated $Q$ dependencies of the intensity of the $M=\pm10 \rightarrow
\pm9$ transition in a single crystal of Mn$_{12}$ for the orientations of the $z$ axis indicated in each
panel. Each panel shows the results for the six exchange-coupling topographies listed in Table~I.}
\end{figure}

Figure~\ref{fig3} shows the experimental results for the intensity of the $M=\pm10 \rightarrow \pm9$
transition in our quasi-single-crystal sample of Mn$_{12}$ for the four orientations with $z$ parallel to
$(1,0,0)$, $(-1,1,0)$, $(0,1,0)$ and $(0,0,1)$, respectively (the vectors refer to the $XYZ$ frame). Clearly,
the $Q$ dependence of the INS intensity depends very markedly on the orientation of the $\textbf{Q}$ vector
with respect to the sample. The calculated $Q$ dependencies are presented in Figure~\ref{fig4} for the six
exchange-coupling topographies listed in Table~I.

Comparison of Figs.~\ref{fig3} and \ref{fig4} reveals that the experimental and theoretical curves exhibit
very similar shapes for each orientation. In particular, the positions of the minima and maxima are in good
agreement. For the $(0,0,1)$ direction, the experimental data shows a maximum at about 0.95~{\AA}$^{-1}$,
which is embraced by two minima at about 0.5 and 1.4~{\AA}$^{-1}$ with a slight upturn towards higher $Q$. In
the $(1,0,0)$ direction, the experiment exhibits a steep drop towards a minimum at about 0.65~{\AA}$^{-1}$,
then a broad maximum at about 1.2~{\AA}$^{-1}$ which is followed by a slow drop at higher $Q$ values. The
$(-1,1,0)$ curve is characterized by a broad asymmetric feature with a maximum at about 0.8~{\AA}$^{-1}$ and
a steeper flank at the side of higher $Q$ values. In the $(0,1,0)$ direction, finally, the experiment shows a
peak at about 1.1~{\AA}$^{-1}$ with indications of two minima at about 0.6 and 1.5~{\AA}$^{-1}$. The
theoretical curves reproduce these features very well.

However, also differences can be noted. In particular, the scattering intensities at the minima are
significantly higher than predicted by theory. In fact, many of the minima are predicted to have almost zero
intensity. This is not unexpected, it is a rather typical observation in non-deuterated samples and due to
the large incoherent scattering of the hydrogen atoms.\cite{Gui04,OW_CsFe8_INS,OW_CsFe8_INS2} Furthermore,
while the relative intensities for the three directions $(1,0,0)$, $(-1,1,0)$, and $(0,1,0)$ are in accord
with the theoretical trend, the $(0,0,1)$ curve is almost a factor of 2 more intense than predicted. This
could be an effect of the sample geometry, which is asymmetric in the sense that a neutron beam along the $z$
axis has to traverse about 1.5~cm of Mn$_{12}$ while a neutron beam in the $xy$ plane traverses at most about
0.8~cm. This suggests that absorption is weaker in the $(0,0,1)$ direction, consistent with the observation.

The theoretical curves in the $(-1,1,0)$ and $(0,1,0)$ directions are basically insensitive to the
exchange-coupling topography, while for the height of the maximum in the $(0,0,1)$ and $(1,0,0)$ directions
the same trend as found for the powder $Q$ dependence, Fig.~\ref{fig2}, is observed. Based on the very weak
intensity expected for the Florentine coupling scheme, which is in contrast with the experiment, this scheme
again is ruled out.

\begin{figure}
\includegraphics{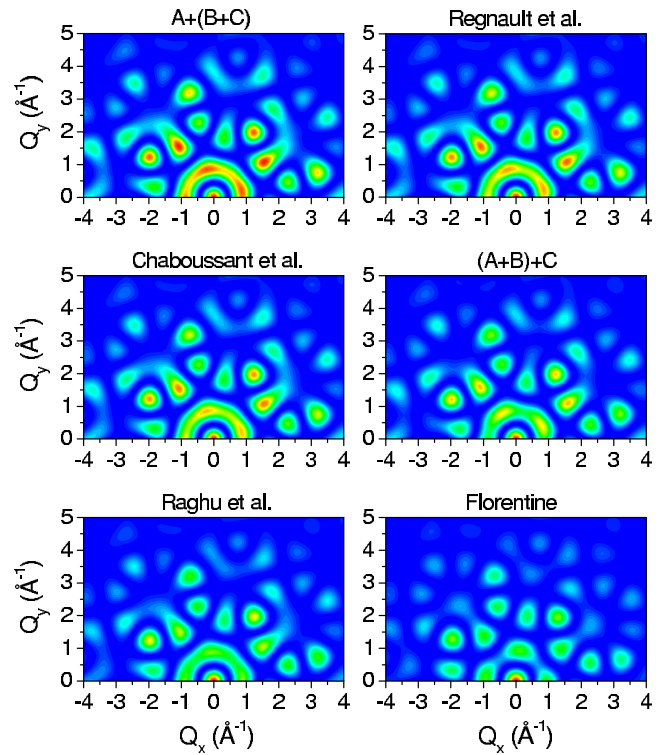}
\caption{\label{fig5}(Color online) Calculated INS intensity of the $M=\pm10 \rightarrow \pm9$ transition in
a single crystal of Mn$_{12}$ for the (0,0,1) orientation as function of the momentum transfer in $x$ and $y$
direction ($Q_z = 0$). The INS intensity is represented by the color [dark gray (blue online) = 0, light gray
(red online) = maximum]. Each panel shows the result for the indicated exchange-coupling topography.}
\end{figure}

Besides the above conclusions we feel that the accuracy of the present experiment is not such as to allow
further conclusions. At this point it is perfectly obvious to ask: how well can the projection coefficients
be retrieved from the $Q$ dependence and what would be an appropriate procedure. In order to explore this,
the calculated INS intensity for the various exchange-coupling topographies for Mn$_{12}$ is shown in
Fig.~\ref{fig5} for the $(0,0,1)$ direction as function of a momentum transfer in the $XY$ plane [the
$(0,0,1)$ curves in the previous Fig.~\ref{fig4} correspond to cuts along the circle $(Q_x-1.065)^2+Q_y^2 =
(0.72)^2$]. Evidently, the single-crystal INS intensity exhibits features of significant intensity at certain
special points, which are related to the molecular structure (\emph{vide infra}). The intensity of these
"towers", however, varies significantly with the exchange-coupling topography. A clear trend of the
intensities is seen along the sequence A+(B+C), Regnault \emph{et al.}, Chaboussant \emph{et al.}, (A+B)+C,
Raghu \emph{et al.}, and Florentine, which is the same trend noted before for the height of the maximum at
about 1~{\AA}$^{-1}$ [see Figs.~\ref{fig2}(a) and \ref{fig4}].

A second, more striking trend can be observed. At $Q \approx 1$~{\AA}$^{-1}$ there is a circle of intensity,
which exhibits four minima and maxima (if one goes around the full 360$^\circ$). While for the (A+B)+C scheme
the left maximum (in the quadrant $Q_y<0$, $Q_x>0$) is at about $-27^\circ$ from the $y$ axis, it rotates to
about $+7^\circ$ in the sequence of (A+B)+C, Regnault \emph{et al.}, Chaboussant \emph{et al.}, A+(B+C),
Raghu \emph{et al.}, and Florentine [the Florentine scheme in fact has its maxima where the (A+B)+C scheme
has its minima, and vice versa]. Observation of the locations of the minima and maxima on the 1~{\AA}$^{-1}$
circle would be hence very interesting.

A qualitative understanding of the intensity pattern can be obtained if one disregards the $Q$ dependence of
the form factors $F_i(Q)$, i.e., assumes $F_i(Q)=1$. The $\cos(\textbf{Q}\cdot\textbf{R}_{ij})$ factors in
$\mathcal{F}(\textbf{Q)}$, Eq.~(\ref{F_S}), can be identically replaced by
$e^{i\textbf{Q}\cdot\textbf{R}_{ij}}$ [the same sum but with $\sin(\textbf{Q}\cdot\textbf{R}_{ij})$ equals
zero]. Then one finds
\begin{eqnarray}
 \mathcal{F}(\textbf{Q}) &=& \int d\textbf{R} e^{i \textbf{Q}\cdot\textbf{R}} \mathcal{G}(\textbf{R}),\\
\label{G_R}
 \mathcal{G}(\textbf{R}) &=& \sum_{ij} \Gamma_1(i)\Gamma_1(j) \delta(\textbf{R}-\textbf{R}_{ij}),
\end{eqnarray}
i.e., $\mathcal{F}(\textbf{Q})$ is just the Fourier transform of $\mathcal{G}(\textbf{R})$. It should be
noted that the INS intensity $I_{nm}(\textbf{Q})$, Eq.~(\ref{I_nm_S}), in general cannot be Fourier
transformed because of the $\textbf{Q}$ dependence of the sum. Exceptions are special cases such as uniaxial
clusters with $\textbf{Q}$ in the $xy$ plane. Then the sum is a constant, see also Eq.~(\ref{I_nm_SMM}), and
also $I_{nm}(\textbf{Q})$ is Fourier transformable [if $F_i(Q)=1$]. Since $\mathcal{F}(\textbf{Q)}$ is the
Fourier transform of the distance vectors $\textbf{R}_{ij}$ (and not of the position vectors $\textbf{R}_i$
as in diffraction), it exhibits maxima at $\textbf{Q}\cdot\textbf{R}_{ij} \approx 2\pi n$, which are,
however, severely broadened since only a small number of metal centers is involved ($n$ is an integer here).

\begin{figure}
\includegraphics{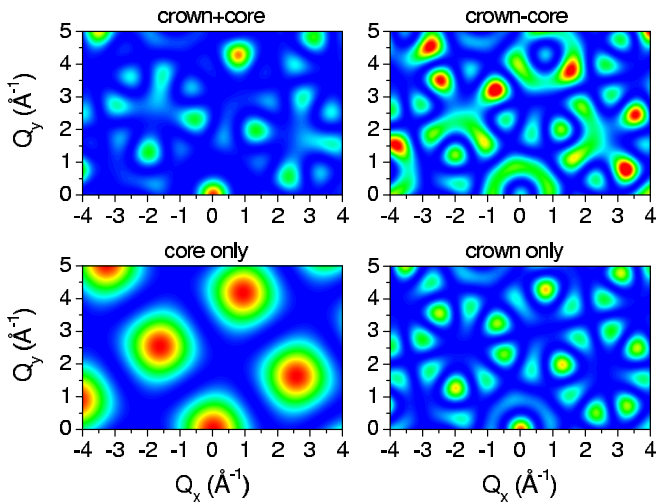}
\caption{\label{fig6}(Color online) $\mathcal{F}(\textbf{Q)}$, Eq.~(\ref{F_S}), of Mn$_{12}$ in the (0,0,1)
orientation as function of the momentum transfer in $x$ and $y$ direction ($Q_z = 0$). The value of
$\mathcal{F}(\textbf{Q)}$ is represented by the color [dark gray (blue online) = 0, light gray (red online) =
maximum]. The panels show the results for the indicated situations discussed in the text.}
\end{figure}

It is revealing to plot $\mathcal{F}(\textbf{Q)}$ for the cases $\{\Gamma_1(A),\Gamma_1(B),\Gamma_1(C)\} =
\{1,1,1\}$ ("crown+core"), $\{-1,1,1\}$ ("crown-core"), $\{0,1,1\}$ ("crown only"), and $\{1,0,0\}$ ("core
only"), as presented in Fig.~\ref{fig6}. The crown+core case represents the direct Fourier transformation of
the "density of distances" $\sum_{ij} \delta(\textbf{R}-\textbf{R}_{ij})$ and mimics a fully ferromagnetic
situation. The crown-core case in contrast mimics an antiferromagnetic alignment of the core spins with
respect to the crown spins. The other two cases focus on the core and crown spins, respectively, i.e.,
neglect the interferences between core and crown.

The core-only plot demonstrates that indeed broadened peaks appear at $\textbf{Q}\cdot\textbf{R}_{ij} \approx
2\pi n$ in $\mathcal{F}(\textbf{Q)}$. The comparison of the crown+core and crown-core plots shows that they
are rather complementary in the sense that at the special $\textbf{Q}$ points the former has maximum
intensity where the latter has minimum intensity, and vice versa. Most interestingly, the intensity circle at
$Q \approx 1$~{\AA}$^{-1}$ is entirely absent in the crown+core plot while it is rather strong in the
crown-core plot. It hence directly reflects the antiferromagnetic spin alignment of the core and crown spins.
The crown-only plot shows the contribution from the interference on the crown. It contributes only very
weakly to the intensity at $Q \approx 1$~{\AA}$^{-1}$.

This discussion shows that each feature visible in one of the plots of Fig.~\ref{fig5} can be related to
certain pairs $(i,j)$ of spin centers, and its amplitude to $\Gamma_1(i)\Gamma_1(j)$. One appropriate
strategy to determine the values of $\Gamma_1(i)\Gamma_1(j)$ would be thus to measure the INS intensity at
the $\textbf{Q}$ points where $\textbf{Q}\cdot\textbf{R}_{ij} = 2\pi n$. Since most features appear for $Q$
$\gtrsim$ 2~{\AA}$^{-1}$, it is now also obvious that the experiment presented in this section, which was
limited to $0.4 < Q < 1.6$~{\AA}$^{-1}$, could not be very successful in discriminating the different
exchange-coupling topographies.

%

\section{Conclusions}

In this work the possibility of extracting the exchange-coupling constants of SMMs from the $Q$ dependence of
the INS intensity of transitions within the ground-state spin multiplet has been explored, with Mn$_{12}$ as
an example. The calculated $Q$ dependence of powder SMM samples was found to show significant variations for
different exchange-coupling situations, but in order to distinguish these experimentally, very precise
measurements are required. The powder $Q$ dependence has a potential to discriminate between very different
exchange-coupling situations. In the case of Mn$_{12}$, the Florentine coupling scheme, which assumes a very
dominating $J_1$ interaction, could be discarded. The $Q$ dependence of single-crystal SMM samples has a
higher potential, because the additional dependence of the INS intensity on the orientation of the momentum
transfer vector $\textbf{Q}$ allows for more independent measurements. The presented experimental results on
a quasi-single-crystal Mn$_{12}$ sample, which consisted of an oriented array of about 500 single crystals,
however, did not allow us to infer information beyond that already concluded from the powder $Q$ dependence.
On the one hand, this is due to the obvious limitations of the experiments. On the other hand, a qualitative
discussion of the situation shows that these experiments did not probe all the relevant $Q$ space and
indicates a better measurement procedure.

As a summary, deducing the exchange-coupling constants in SMMs from the INS $Q$ dependence seems to be a
viable strategy, but is experimentally challenging. It requires precise work in order to arrive at
quantitative conclusions concerning the coupling constants. For instance, highly deuterated samples seem to
be mandatory. Also, absorption corrections due to the actual sample geometry probably have to be employed,
very much as in X-ray crystallography. However, although these issues clearly need further development, they
are within reach of today's spectrometers.

The key point in the approach is the relationship between the exchange-coupling constants and the spin
density map. For the SMMs it was noted in this work that the spin density map can be determined from the $Q$
dependence of the INS transitions within the spin ground state of the SMM. However, a similar approach is
clearly also possible with other experimental techniques, which allow one to measure the spin density map,
such as elastic neutron scattering or nuclear magnetic resonance.

%

\begin{acknowledgments}
Financial support by the Swiss National Science Foundation and the European Union (EC-RTN-QUEMOLNA, Contract
No. MRTN-CT-2003-504880) is gratefully acknowledged.
\end{acknowledgments}

%

\appendix
\section{}

In this Appendix, a relationship between the INS formulas for single-crystal and powder SMMs, i.e.,
Eqs.~(\ref{I_nm_S}),(\ref{F_S}) and (\ref{I_nm_powder2_S}),(\ref{F_powder S}), respectively, shall be
discussed. As mentioned already in Sec.~VI, the interference factor $\mathcal{F}({\bf Q})$ of a
single-crystal SMM, Eq.~(\ref{F_S}), can be written identically with $\cos({\bf Q}\cdot{\bf R}_{ij})$
replaced by $e^{i{\bf Q}\cdot{\bf R}_{ij}}$. Expanding the exponential in spherical harmonics,
$\mathcal{F}({\bf Q})$ can be rewritten as
\begin{equation}
\label{eqA1}
 \mathcal{F}({\bf Q}) = 4 \pi \sum_{kq} (i)^k (-1)^q Y^{(k)}_q({\bf Q}) C_{k} \mathcal{F}^{kq}(Q),
\end{equation}
where
\begin{eqnarray}
\label{eqA2}
 \mathcal{F}^{kq}(Q) &=& \sum_{ij} f^{kq}(Q,{\bf R}_{ij}) \Gamma_1(i) \Gamma_1(j),\\
\label{eqA3} f^{kq}(Q,\textbf{R}_{ij}) &=& F_i(Q) F_j(Q) j_k(QR_{ij}) T^{(k)}_q({\bf R}_{ij}).\qquad
\end{eqnarray}
Here, $k$ may be restricted to even values, $k=0,2,4,\ldots$. $Y^{(k)}_q({\bf V})$ denotes the $q$th
component of the spherical harmonics of rank $k$ related to the Cartesian vector $\textbf{V}$. It is
proportional to the irreducible spherical tensor $T^{(k)}_q({\bf V})$, $Y^{(k)}_q({\bf V}) = C_{k}
T^{(k)}_q({\bf V})$, which also defines the proportionality constant $C_{kq}$.

This is the main finding of this Appendix: Equations~(\ref{eqA2}) and (\ref{eqA3}) are identical to the
results for the interference factors of powder SMM samples $-$ if one expands $I_{nm}(Q)$ in terms of the
spherical tensors $T^{(k)}_q$ and not the symmetrized spherical tensors $U^{(k)}_q$ as in
Eq.~(\ref{I_nm_powder2_S}).

This point deserves comment. Equations~(\ref{eqA3}) and (\ref{F_powder S}) look exactly identical, but in
fact differ in the definition of the interference factors $f^{kq}(Q,\textbf{R}_{ij})$:
Equation~(\ref{F_powder S}) is derived from Eq.~(\ref{I_nm_powder2}), which in turn is obtained by rewriting
the general INS formula for powder samples, Eq.~(\ref{I_nm_powder}), in terms of the symmetrized spherical
tensors $U^{(k)}_q$. Hence, the expressions for the $f^{kq}(Q,\textbf{R}_{ij})$ contain the tensors
$U^{(k)}_q({\bf R}_{ij})$.\cite{OW_INS2} An expansion in terms of $U^{(k)}_q$ is most convenient for explicit
calculations. The general powder formula Eq.~(\ref{I_nm_powder}), however, also can be expressed in terms of
the spherical tensors $T^{(k)}_q$. The resulting equations look exactly the same as that for $U^{(k)}_q$, but
with all $U^{(k)}_q$ replaced by $T^{(k)}_q$ [for instance, replacing $T^{(k)}_q$ by $U^{(k)}_q$ in
Eq.~(\ref{eqA3}) directly yields the equations (B2) of Ref.~[\onlinecite{OW_INS2}]]. In this Appendix, the
$T^{(k)}_q$-expansion is preferred because the algebra is then more obvious, but in the expressions all
$T^{(k)}_q$ can be replaced identically by $U^{(k)}_q$. In this sense, Eqs.~(\ref{eqA2}) and (\ref{eqA3}) are
identical to the results for powder SMM samples, such that the above equations establish the sought-after
relationship.

The relation Eq.~(\ref{eqA1}) is convenient. It separates the dependence of the INS intensity on the
orientation of the $\textbf{Q}$ vector from the dependence on the magnitude $Q$ and the cluster properties
[which enter via $\textbf{R}_{ij}$ and $\Gamma_1(i)$]. Furthermore, it establishes the connection between the
single-crystal and powder interference factors (for the powder INS intensity, however, only the factors with
$k=0,2$ are relevant, while in the single-crystal case no such restriction exists).

Using the orthogonality relation for spherical harmonics, one obtains
\begin{equation}
\label{eqA4}
 \int \frac{d\Omega}{4 \pi} \mathcal{F}({\bf Q}) Y^{(k)}_q({\bf Q}) =  (i)^k C_{k} F^{kq}(Q).
\end{equation}
Hence, integrating the interference factor $\mathcal{F}({\bf Q})$ over all orientations of $\textbf{Q}$ for a
given magnitude $Q$ allows one to extract the interference factors $F^{kq}(Q)$.

If one assumes $F_i(Q)= 1$ for the moment, and uses the closure relation of spherical Bessel functions, one
obtains
\begin{eqnarray}
\label{eqA5}
 \int dQ \: Q^2 \mathcal{F}^{(k)}_q(Q) j_k(Q R) =\cr
 \cr \frac{\pi}{2R^2} \sum_{ij} \Gamma_1(i) \Gamma_1(j) T^{(k)}_q({\bf R}_{ij}) \delta( R-R_{ij}),
\end{eqnarray}
where $R$ is an arbitrary number [the result also holds for the weaker condition of $F_i(Q)= F(Q)$ for all
$i$ if $\mathcal{F}^{kq}(Q)/F(Q)$ is considered instead of $\mathcal{F}^{kq}(Q)$]. Hence, integrating
$\mathcal{F}^{kq}(Q)$ over $Q$ using a value of $R$ which matches one of the metal distances $R_{ij}$ allows
one to project out the value of $\Gamma_1(i) \Gamma_1(j)$ for the pair ($i$,$j$) [to be precise, one has to
assume that if $R_{ij}$ is equivalent for several pairs ($i$,$j$) that then also $\Gamma_1(i) \Gamma_1(j)$ is
equivalent for these pairs, which is reasonable]. If one combines Eqs.~(\ref{eqA4}) and (\ref{eqA5}) and sums
over $\sum_{kq} T^{(k)*}_q({\bf R}_{ij})$, then one recovers the Fourier transform of $\mathcal{F}({\bf Q})$,
Eq.~(\ref{G_R}).

A final comment: Obviously, very similar mathematics can be performed in order to establish relations between
the INS formulas Eq.~(\ref{I_nm}) and Eq.~(\ref{I_nm_powder2}), which are valid for general spin clusters.
Following the above lines, the corresponding equations are simple to obtain. They hence shall not be
reproduced here.

%

%

\begin{references}

\bibitem{Ses93n}
R. Sessoli, D. Gatteschi, A. Caneschi, and M. A. Novak, Nature (London) {\bf 365}, 141 (1993).

\bibitem{Fri96}
J. R. Friedman, M. P. Sarachik, J. Tejeda, and R. Ziolo, Phys. Rev. Lett. {\bf 76}, 3830 (1996).

\bibitem{Tho96}
L. Thomas, F. Lionti, R. Ballou, D. Gatteschi, R. Sessoli, and B. Barbara, Nature (London) {\bf 383}, 145
(1996).

\bibitem{Mn12_Fe8}
G. Christou, D. Gatteschi, D. N. Hendrickson, and R. Sessoli, MRS Bull. \textbf{25}, 66 (2000); D. Gatteschi
and R. Sessoli, Angew. Chem. Int. Ed. {\bf 42}, 268 (2003).

\bibitem{Ben90}
A. Bencini and D. Gatteschi, {\it Electron Paramagnetic Resonance of Exchange Coupled Cluters} (Springer,
Berlin, 1990).

\bibitem{Liv02}
E. Liviotti, S. Carretta, and G. Amoretti, J. Chem. Phys. \textbf{117}, 3361 (2002).

\bibitem{OW_INS2}
O. Waldmann and H. U. G\"udel, Phys. Rev. B \textbf{72}, 094422 (2005).

\bibitem{Hil06}
A. Wilson, J. Lawrence, E.-C. Yang, M. Nakano, D. N. Hendrickson, and S. Hill, Phys. Rev. B \textbf{74},
140403 (2006).

\bibitem{Acc06}
S. Accorsi, A.-L. Barra, A. Caneschi, G. Chastanet, A. Cornia, A. C. Fabretti, D. Gatteschi, C. Mortalo, E.
Olivieri, F. Parenti, P. Rosa, R. Sessoli, L. Sorace, W. Wernsdorfer, L. Zobbi, J. Am. Chem. Soc.
\textbf{128}, 4742 (2006).

\bibitem{Bir04}
R. Bircher, G. Chaboussant, A. Sieber, H. U. G\"udel, and H. Mutka, Phys. Rev. B \textbf{70}, 212413 (2004).

\bibitem{Mn12_E}
Despite the tetragonal crystal structure of Mn$_{12}$, also a term $E(\hat{S}_x^2-\hat{S}_y^2)$ has been
found to be significant, see e.g. Refs.~[\onlinecite{Chu01,Hil01,Cor02,Bir04}]]. This term affects strongly
the tunneling splittings, but not the wave functions and anisotropy splitting of the states of relevance in
this work. As the tunneling splittings are way to small to be detected in INS experiments, the actual terms
responsible for tunneling can be disregarded for the purposes of this work.

\bibitem{Chu01}
E. M. Chudnovsky and D. A. Garanin, Phys. Rev. Lett. \textbf{87}, 7203 (2001).

\bibitem{Hil01}
S. Hill, R. S. Edwards, S. I. Jones, N. S. Dalal, and J. M. North, Phys. Rev. Lett. \textbf{90}, 217204
(2003).

\bibitem{Cor02}
A. Cornia, R. Sessoli, L. Sorace, D. Gatteschi, A. L. Barra, and C. Daiguebonne, Phys. Rev. Lett.
\textbf{89}, 257201 (2002).


\bibitem{Ste63}
K. W. H. Stevens, in \emph{Magnetism}, edited by G. T, Rado and H. Suhl (Academic Press, New York, 1963),
Vol. I.

\bibitem{Cac98}
R. Caciuffo, G. Amoretti, A. Murani, R. Sessoli, A. Caneschi, and D. Gatteschi, Phys. Rev. Lett. \textbf{81},
4744 (1998).

\bibitem{Mir99}
I. Mirebeau, M. Hennion, H. Casalta, H. Andres, H. U. G\"udel, A. V. Irodova, and A. Caneschi, Phys. Rev.
Lett. \textbf{83}, 628 (1999).

\bibitem{Fur77}
A. Furrer and H. U. G\"udel, Phys. Rev. Lett. \textbf{39}, 657 (1977).

\bibitem{OW_INS}
O. Waldmann, Phys. Rev. B \textbf{68}, 174406 (2003).

\bibitem{Hen97}
M. Hennion, L. Pardi, I. Mirebeau, E. Suard, R. Sessoli, and A. Caneschi, Phys. Rev. B \textbf{56}, 8819
(1997).



\bibitem{Fur79}
A. Furrer and H. U. G\"udel, J. Magn. Magn. Mater. \textbf{14}, 256 (1979).

\bibitem{Gue85}
H. U. G\"udel, in \emph{Magneto-Structural Correlations in Exchange-Coupled Systems}, edited by R. D. Willet
(Reidel, Amsterdam, 1985), p. 325.

\bibitem{OW_CsFe8_INS2}
O. Waldmann, C. Dobe, H. U. G\"udel, and H. Mutka, Phys. Rev. B \textbf{74}, 054429 (2006).



\bibitem{Lis80}
T. Lis, Acta Crystallogr., Sect. B \textbf{35}, 2042 (1980).

\bibitem{Ses93}
R. Sessoli, H.-L. Tsai, A. R. Schake, S. Wang, J. B. Vincent, K. Folting, D. Gatteschi, G. Christou, and D.
N. Hendrickson, J. Am. Chem. Soc. \textbf{115}, 1804 (1993).

\bibitem{Har96}
F. Hartmann-Boutron, P. Politi, and J. Villain, Int. J. Mod. Phys. B \textbf{10}, 2577 (1996).

\bibitem{Rag01}
C. Raghu, I. Rudra, D. Sen, and S. Ramasesha, Phys. Rev. B \textbf{64}, 064419 (2001).

\bibitem{Reg02}
N. Regnault, Th. Jolicoeur, R. Sessoli, D. Gatteschi, and M. Verdaguer, Phys. Rev. B \textbf{66}, 054409
(2002).

\bibitem{OW_SPINDYN}
O. Waldmann, Phys. Rev. B \textbf{65}, 024424 (2002).

\bibitem{Cha04} G. Chaboussant, A. Sieber, S. Ochsenbein, H.-U. G\"udel, M. Murrie, A. Honecker,
N. Fukushima, and B. Normand, Phys. Rev. B \textbf{70}, 104422 (2004).

\bibitem{Hon05}
A. Honecker, N. Fukushima, B. Normand, G. Chaboussant, and H.-U. G\"udel, J. Magn. Magn. Matter.
\textbf{290}, 966 (2005).

\bibitem{Bou02}
D. W. Boukhvalov, A. I. Lichtenstein, V. V. Dobrovitski, M. I. Katsnelson, B. N. Harmon, V. V. Mazurenko, and
V. I. Anisimov, Phys. Rev. B \textbf{65}, 184435 (2002).

\bibitem{Par04}
K. Park, M. R. Pederson, and C. S. Hellberg, Phys. Rev. B \textbf{69}, 014416 (2004).

\bibitem{Kat99}
M. I. Katsnelson, V. V. Dobrovotski, and B. N. Harmon, Phys. Rev. B \textbf{59}, 6919 (1999).

\bibitem{Zve96}
A. K. Zvezdin and A. I. Popov, Sov. Phys. JETP \textbf{82}, 1140 (1996).

\bibitem{Tup02}
I. Tupitsyn and B. Barbara, in \emph{Magnetism: Molecules to Materials. Nanosized Magnetic Materials}, edited
by J. S. Miller and M. Drillon (Wiley-VCH, Weinheim, 2002, p. 109.



\bibitem{OW_TRINS}
O. Waldmann, G. Carver, C. Dobe, D. Biner, A. Sieber, H. U. G\"udel, H. Mutka, J. Ollivier, N. E. Chakov,
Appl. Phys. Lett. \textbf{88}, 042507 (2006).




\bibitem{FFILL}
P. J. Brown, in {\it Neutron Data Booklet}, edited by A. J. Dianoux and G. Lander (Institute Laue-Langevin,
Grenoble, 2001).



\bibitem{lamp}
LAMP, the Large Array Manipulation Program. http://www.ill.fr/data\_treat/lamp/front.html.

\bibitem{Gui04}
T. Guidi, S. Carretta, P. Santini, E. Liviotti, N. Magnani, C. Mondelli, O. Waldmann, L. K. Thompson, and L.
Zhao, Phys. Rev. B \textbf{69}, 104432 (2004).

\bibitem{OW_CsFe8_INS}
O. Waldmann, C. Dobe, H. Mutka, A. Furrer, and H. U. G\"udel, Phys. Rev. Lett. \textbf{95}, 057202 (2005).


\end{references}
\end{document}